\documentclass[preprint,nofootinbib,nobibnotes,groupaddress,preprintnumbers,aps,prd]{revtex4}
\usepackage{tipa}
\usepackage{bm,epsfig,mathrsfs,amsmath,amssymb,graphicx}
\usepackage{epsfig,amsmath,graphicx,amssymb}
\usepackage{graphicx}
\usepackage{dcolumn}
\usepackage{bm}
\usepackage[toc,title,titletoc,header]{appendix}
\def\dbar{{\mathchar'26\mkern-12mu d}}

\date{\today}

\begin{document}

\title{Quantum refrigerators in finite-time cycle duration}
\author{Guangqian Jiao$^{1}$}
\author{Shoubao Zhu$^1$}

\author{Jizhou He$^{1}$}
\author{Yongli Ma$^{2}$}
\author{Jianhui Wang$^{1,2}$}\email{wangjianhui@ncu.edu.cn}

\affiliation{ $^1\,$ Department of Physics, Nanchang University,
Nanchang 330031, China\\ $^2\,$   State Key Laboratory of Surface
Physics and Department of Physics, Fudan University, Shanghai
200433, China}

\begin{abstract}
We derive cooling rate and coefficient of performance  as well as
their variances  for a quantum Otto engine proceeding in finite-time
cycle period. This machine consists of two driven strokes, where the
system isolated from the heat reservoir undergoes finite-time
unitary transformation, and two isochoric steps, where the
finite-time system-bath interaction durations take the system away
from the equilibrium even at the ends of the two stages. We
explicitly calculate the  statistics of cooling rate and coefficient
of performance for the machine  operating with  an analytically
solvable two-level system. We clarify the role of finite-time
durations of four processes on the machine performance. We  show
that there is the trade-off between the performance parameter and
its corresponding variance, thereby indicating that
 the cooling rate or coefficient of performance  can be enhanced, but at
 the cost of increasing the corresponding fluctuations.

 PACS number(s): 05.70.Ln

\end{abstract}

\maketitle
\date{\today}
\section{Introduction}
A refrigerator as an inverse operation of a heat engine transfers
energy from a cold thermal bath of temperature $T_h$ to a a hot one
with temperature $T_c$ by consuming work. Quantum refrigerators well
as quantum heat engines use quantum systems as their working
substance and can be classified either cyclic \cite{Aba12, Su16,
Fel00, Quan07, Ass19, Hong20, Rez06} or steady-state \cite{Kos14,
Cle12, Mas19, Ton05, Lev12, Lin10} models. The quantum Otto cycle of
operation, as a typical example of cyclic machines,  is controlled
by the segments of time that the working system is coupled to a hot
and a cold bath, and by the time interval required to driving the
control parameter of the system. It was most studied \cite{Aba12,
Su16, Fel00,  Ass19, Hong20, Rez06, Ale15, Lee20} as it is easier to
analyze and realize. To describe the performance characteristics of
a refrigerator, one introduces the coefficient of performance (COP)
that is defined as the ratio of heat absorbed from the cold
reservoir and work input. An upper bound on the COP imposed by the
second law of thermodynamic is given by the Carnot COP:
$\varepsilon_C=T_c/(T_h-T_c)$, which, however, requires
infinitesimally slow transitions between thermodynamic states and
thus produces vanishing cooling rate. Hence, the refrigerators
actually operate far from the infinite long time limit in order for
positive cooling rate to be produced \cite{Wang12, Tom12, Hu13,
Izu15}. The finite cooling rate for a cyclic refrigerator consisting
of a sequence of thermodynamic processes indicates that each process
must proceed in finite time.  For an adequate description of an
actual machine, the effects induced by finite-time duration along
any thermodynamic stroke on heat and work have to be considered.

While in a macroscopic  system  the work and heat are deterministic,
for a microscopic quantum
 system (with a limited number of freedoms) these physical
variables become random due to non-negligible thermal \cite{Sei12,
Sek10} and quantum \cite{Esp09, Cam11} fluctuations. Theoretical and
experimental investigation on the statistics of work \cite{Esp14,
Quan20, Liu19, Sol13, Hek13, Bat14, Cer17} and heat \cite{Rah12,
Den18, Gas14} has attracted much interest in the literature. On the
other hand, for heat engines the statistics of
 power \cite{Hong20, Hol18, Hol14, Wang19} and efficiency \cite{Lutz20, Vro16, Jiang15, Vro20, Ver14, Ver15, Par16}    has been analyzed,
 under the assumption that either system-bath interaction interval
  or unitary driving process is quasistatic.  However, a unified
thermodynamic description of quantum refrigerators with
non-negligible quantum and thermal fluctuations, particularly when
every thermodynamic stroke of these machines evolves in finite time,
is available.

In the present paper, we study the thermodynamics of a quantum Otto
refrigerator where all the four strokes proceed in finite time,
within a framework of stochastic thermodynamics. Having determined
distribution functions for heat and work, we derive general formulae
for the
 COP and cooling load as well as their variances. With these we then
analyze a quantum Otto cycle working with a two-level system which
is exactly solvable analytically. We discuss the effects of thermal
and quantum fluctuations on finite-time performance and the
statistics of the machine, and also demonstrate that there is
trade-off of the physical variable (COP or cooling rate) and its
fluctuations. We finally show that, the average COP $\langle
\varepsilon\rangle$ can be always larger than the conventional
thermodynamic COP $\varepsilon_{th}$ for adiabatic driving, but it
can be equal to or smaller than COP $\varepsilon_{th}$ for
nonadiabatic driving.

\section{The probability of stochastic COP for   quantum otto refrigerators}

\begin{figure}[tb]
\includegraphics[width=2.8in]{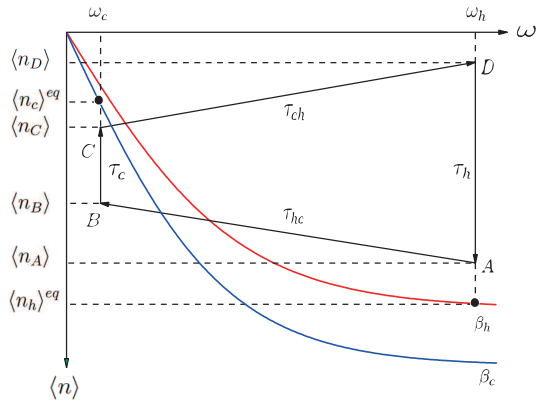}
 \caption{(Color online) Schematic diagram of a quantum Otto  refrigeration cycle operating
with a two-level system in $(\omega, n)$ plane. The cycle consists
of  the two adiabatic strokes (connecting states $A$ and $B$, and
$C$ and $D$), where the system isolated from the heat reservoir
evolves unitary transformation, and two isochoric strokes
(connecting states $B$ and $C$, and $D$ and $A$), where the system
is kept in thermal contact with  the cold and the hot reservoir,
respectively. The average population $n_3$ ($n_1$)  at the end of
the cold (hot) isochoric stroke would approach its asymptotic value
$n_c^{eq}$ ($n_h^{eq}$), when and only  when the system-bath
interaction duration $\tau_{c}$ ($\tau_h$) tends to be infinitely
long. }\label{model}
\end{figure}
The quantum Otto cycle under consideration is sketched in Fig.
\ref{model}. It consists of two isochoric branches, one with a cold
and another with a hot heat reservoir where the Hamiltonian is kept
constant,  and two other strokes, where the system undergoes unitary
transformation while isolating from the thermal reservoirs.
 In
the adiabatic branch $A\rightarrow B$, the system is isolated from
any heat reservoir and undergoes a unitary expansion from time $t=0$
to $t=\tau_{hc}$. Initially, the system is assumed to be with local
thermal equilibrium at inverse temperature $\beta_A(\ge\beta_h)$.
The probability distribution of stochastic work done by the system,
$w_{hc}$, can be given by \cite{Tal07}
\begin{equation}
p(w_{hc})=\sum_{n,m}\delta[w_{hc}-(E_{m}^{c}-E_{n}^{h})]p_{n\rightarrow
m}^{\tau_{hc}}p_{n}^{0}(\beta_A).     \label{pw0a}
\end{equation}
where $E_{n}^{h}$ and $E_{m}^{c}$ are the respective energy
eigenvalues at the initial and final instants along this expansion.
Here
 $p_{n}^{0}(\beta_A)=\frac{e^{-\beta_{A}E^{h}_{n}}}{Z_{A}}$, with the partition funciton  $Z_{A}=\sum_n e^{-\beta E_n}$,
 denotes  the
thermal occupation probability at instant $A$, and
 $p_{n\rightarrow m}^{\tau_{hc}}=|\langle n|U_{\mathrm{exp}}|m\rangle|^2$
is  the transition probability from eigenstate $|n\rangle$ and
$|m\rangle$,  with  the unitary time evolution operator
$U_{\mathrm{exp}}$.   and the population $\langle n\rangle$ remains
unchanged in the adiabatic stage ($\xi=0$).  If the system
Hamiltonian  evolves slowly enough wherein  the quantum adiabatic
condition \cite{Born28} is satisfied  in time interval $\tau_{hc}$,
the the system remains in the same state and the transition
probability therefore satisfies $p_{n\rightarrow
m}^{\tau_{hc}}=\delta_{nm}$, with the Dirac's delta function
$\delta$.

In the next step $B\rightarrow C$, the quantum system with constant
frequency $\omega=\omega_c$ is kept in contact with   a cold thermal
bath of inverse temperature $\beta_c$ during a period $\tau_c$. The
probability density of the stochastic heat $q_c$ can be determined
by conditional distribution to arrive at
\begin{equation}
p(q_c|w_{hc})=\sum_{k,l}\delta[q_c-(E_{l}^{c}-E_{k}^{c})]p_{k\rightarrow
l}^{\tau_c}p_{k}^{\tau_{hc}} ,     \label{pqpk}
\end{equation}
where  $p_{k}^{\tau_{hc}}$ is the occupation probability at time
$t=\tau_{hc}$ and it satisfies the constraint
$p_{k}^{\tau_{hc}}=\delta_{km}$. We assume that at the end of  the
system-bath interaction interval, the system is at the local thermal
equilibrium state with inverse temperature $\beta_C(\ge\beta_c^r)$,
thereby indicating that $p_{k\rightarrow
l}^{\tau_c}={e^{-\beta_{C}E^{c}_{l}}}/{Z_{C}}$ with partition
function $Z_{C}=\sum_l {e^{-\beta_{C}E^{c}_{l}}}$. Without loss of
generality, the internal energy of system  along the isochoric
process can be expressed as $\langle n(t)\rangle=[Z(t)]^{-1}\sum_n n
e^{-\beta(t) E_n^c},$ where $\langle n\rangle$ is the population
(which is also the the expectation value of the particle number
operator in Appendix \ref{tip}). When the time duration $\tau_c$
tends to be infinity, the mean population $n_C$ at end of the
isochoric process approaches the equilibrium value,
\begin{equation}
\langle n_c\rangle^{eq}\equiv
n(\tau_c\rightarrow\infty\rangle=Z^{-1}\sum_n n e^{-\beta_cE_n},
\label{ncn}
\end{equation}
with $Z_c=\sum_n n e^{-\beta_cE_n}$.  It is shown in Appendix
\ref{tip} that the dynamics of the system along the thermalization
can be described by the motion equation  of population, which gives
\begin{equation}
\langle n_c\rangle^{eq}=\langle n_c\rangle^{eq}+[\langle
n_B\rangle-\langle n_c\rangle^{eq}]e^{-\gamma_c\tau_c}, \label{ncc}
\end{equation}
where $\langle n_C\rangle=\langle n(\tau_{hc}+\tau_{c})\rangle$ and
$\langle n_B\rangle=\langle n(\tau_{hc})\rangle$, and $\gamma_c$ is
the thermal conductivity when $\omega=\omega_c$.

 On the
adiabatic compression $C\rightarrow D$, the system is isolated in
time duration $\tau_{ch}$ while the energy gap $\omega$ varies from
$\omega_c$ to $\omega_h$. For given work output $w_{hc}$ and the
heat $q_c$ released from the system, the probability density of
stochastic work input is given by
\begin{equation}
p(w_{ch}|q_c,w_{hc})=\sum\delta[w_{ch}-(E_{j}^{h}-E_{i}^{c})]p_{i\rightarrow
j}^{\tau_{ch}}p_{i}^{\tau_{hc}+\tau_c} ,     \label{pwi2}
\end{equation}
where the occupation probability $p_{i}^{\tau_{hc}+\tau_c}=\delta_{il}$,
and  $p_{i\rightarrow j}^{\tau_{ch}}=|\langle
i|U_{\mathrm{com}}|j\rangle|^2$ is transition probability from
eigenstate $|i\rangle$ and $|j\rangle$, with the time evolution
operator $U_{\mathrm{com}}$ along the compression.

In the fourth step $D\rightarrow A$, the system is coupled to a hot
reservoir of inverse temperature $\beta_h$ in time duration
$\tau_{h}$ while keeping its frequency in a constant with
$\omega=\omega_h$. Since the system returns to its initial state $A$
after the cycle period
$\tau_{cycle}=\tau_{hc}+\tau_c+\tau_{ch}+\tau_h$, we will do not
derive the expression of the stochastic heat exchanged $q_h$ along
this process. As shown in Appendix \ref{tip}, the populations at the
beginning and end of the heating process ($\langle n_D\rangle$ and
$\langle n_A\rangle$) satisfies the constraint:
\begin{equation}
\langle n_A\rangle=\langle n_h\rangle^{eq}+[\langle
n_D\rangle-\langle n_h\rangle^{eq}]e^{-\gamma_h\tau_h}, \label{nah}
\end{equation}
where $\langle n_A\rangle=\langle (\tau_{cycle}), \langle
n_D\langle=\langle n (\tau_{cycle}-\tau_h)\rangle$,
\begin{equation}
\langle n_h\rangle^{eq}\equiv\langle
n(\tau_h\rightarrow\infty)\rangle^{eq}=Z_h^{-1}\sum_n n e^{-\beta_h
E_n} \label{nhn}
\end{equation}
with $Z_h=\sum_ne^{-\beta_h E_n}$, and $\gamma_h$ is the thermal
conductivity between the system and the hot reservoir.


The  probability $p(w_{ch},q_c,w_{hc})$ for the machine which has
certain values of $w_{ch},q_c,w_{hc}$ can be calculated from the the
chain rule for condition probabilities  $p(w_{ch},q_c,w_{hc}) =
p(w_{ch}|q_c,w_{hc})p(q_c\big|w_{hc})p(w_{hc})$:
\begin{equation}
\begin{split}
p(w_{ch},q_c,w_{hc})=&\sum\delta\left[q_c-\left(E_{k}^{c}-E_{m}^{c}\right)\right]\delta\left[w_{hc}-\left(E_{m}^{c}-
E_{n}^{h}\right)\right]\delta\left[w_{ch}-\left(E_{l}^{h}-E_{k}^{c}\right)\right]\\
&\times|\langle n|U_{\mathrm{exp}}|m\rangle|^2|\langle
i|U_{\mathrm{com}}|j\rangle|^2
\frac{e^{-\beta_{A}E^{h}_{n}}e^{-\beta_{C}E^{c}_{k}}}{Z_{A}Z_{C}}.
\end{split}  \label{pwzc}
\end{equation}
In deriving this, we have used  Eqs. (\ref{pw0a}), (\ref{pqpk}) and
(\ref{pwi2}). For the quantum Otto refrigerator,  the stochastic
coefficient of performance reads
$\varepsilon={q_c}/({w_{hc}+w_{ch}})$. It follows, integrating over
all values of $w_{ch},q_c,w_{hc}$, that  the probability
distribution $p(\varepsilon)$   becomes
\begin{equation}
p\left(\varepsilon\right)=\sum_{m,n,i,j}\delta\left(\epsilon-\frac{E_{k}^{c}-E_{m}^{c}}{E_{m}^{c}-E_{n}^{h}+E_{l}^{h}-E_{k}^{c}}\right)
\times\frac{e^{-\beta_{A}E^{h}_{n}}e^{-\beta_{C}E^{c}_{k}}}{Z_{A}Z_{C}}
|\langle n|U_{\mathrm{exp}}|m\rangle|^2|\langle
i|U_{\mathrm{com}}|j\rangle|^2.     \label{p1}
\end{equation}
While for adiabatic driving, the system remains in the same state
($n=m, i=j$), in the nonadiabatic driving the transition probability
$|\langle n|U_{\mathrm{exp}}|m\rangle|^2$ (or $|\langle
i|U_{\mathrm{com}}|j\rangle|^2)$ is positive due to transitions
between sates $n$ and $m$ (or $i$ and $j$). For the quantum Otto
refrigerator,  its efficiency statistics  is fully determined by the
unitary time evolution  for the adiabatic expansion and compression
($U_{\mathrm{exp}}$ and $U_{\mathrm{com}}$), and by the finite-time
system dynamics along the thermalization processes, when the two
temperatures of the two heat reservoirs ($\beta_h$ and $\beta_c$)
are given.
\section{a quantum Otto refrigerator using a two-level system}
We now consider a quantum Otto refrigerator operating with   a
two-level system of the  eigenenergies $E_+=-\hbar\omega/2$ and
$E_-=-\hbar\omega/2$. If the unitary expansion and compression
during the Otto cycle (from $\omega_h$ to $\omega_c$ to $\omega_h$
and vice versa) is such that there is a probability of level
transitions due to quantum fluctuations,  then there is a
probability that population $\langle n\rangle$ may change with
varying time. After a simple calculation (see Appendix \ref{rep}),
we find that
\begin{equation}
 \langle n_B\rangle=(1-2\xi)\langle n_A\rangle, ~ \langle n_D\rangle=(1-2\xi)\langle n_C\rangle \label{nbnc}
\end{equation}
where   $\xi=|\langle \pm|U_{\mathrm{exp}}|\mp\rangle|^2=|\langle
\pm|U_{\mathrm{com}}|\mp\rangle|^2$ is called the adiabacity
parameter indicating the probability of transition between state
$|+\rangle$ and $|-\rangle$ during the compression or expansion. As
shown in Fig. \ref{model}, the populations at any instant along the
cycle is negative, which means that $\xi$ must be situated between
$0\le\xi<1/2$. The probability of no state transition along either
driving phase is accordingly $|\langle
\pm|U_{\mathrm{exp}}|\pm\rangle|^2=|\langle
\pm|U_{\mathrm{com}}|\pm\rangle|^2=1-\xi$. The adiabaticity
parameter $\xi$ depends on the the speed at which the driving
process is performed \cite{Aba12, Ass19, Lutz20}. When the time
scale of the state change is much larger than that of the dynamical
one, the quantum adiabatic condition is satisfied  and the
population $\langle n\rangle$ remains unchanged in the adiabatic
stage ($\xi=0$). Rapid change in the control field $\omega$,
however, leads to nonadiabatic behavior ($\xi>0$) which
 can understood as inner friction \cite{Rez06, Fel00, Ale15, Cor15, Cam19, Pla14} causing state transitions. Equation (\ref{nbnc}) shows that
 $\langle n_C\rangle> \langle n_B\rangle$ and $\langle n_D\rangle>\langle n_C\rangle$ for $\xi>0$
(see also Fig. \ref{model}), as lies in the fact that the finite
time duration of the expansion and compression accounts for the
nonadiabatic inner friction related to the irreversible entropy
production.

Using Eqs. (\ref{ncc}), (\ref{nah}), and (\ref{nbnc}), it follows
that the populations $\langle n_A\rangle$ and $\langle n_C\rangle$
can be expressed in terms of the equilibrium populations $\langle
n_h\rangle^{eq}$ and $\langle n_c\rangle^{eq}$,
\begin{equation}
\langle n_A\rangle=\langle n_h \rangle^{eq}+\Delta_h, \langle
n_C\rangle=\langle n_c \rangle^{eq}+\Delta_c
 \end{equation}
where
\begin{equation}
\Delta_h=\frac{\left(2\xi-1\right)\left[\left(2\xi-1\right)\langle
n_h\rangle+\langle n_c\rangle^{eq}\right] -y\left[\langle
n_h\rangle+\left(2\xi-1\right)\langle n_c\rangle^{eq}\right]}
{xy-\left(2\xi-1\right)^2},
\end{equation}
\begin{equation}
\Delta_c=\frac{\left(2\xi-1\right)\left[\left(2\xi-1\right)\langle
n_c\rangle+\langle n_h\rangle^{eq}\right] -x\left[\langle
n_c\rangle+\left(2\xi-1\right)\langle n_h\rangle^{eq}\right]}
{xy-\left(2\xi-1\right)^2},
\end{equation}
with  $x=e^{\gamma_h\tau_h}$ and $y=e^{\gamma_c\tau_c}$. Hereafter
we will refer $x$ and $y$ rather than $\tau_h$ and $\tau_c$ as the
time durations along the hot and cold isochoric branches for
simplicity. Here $\langle n_c\rangle^{eq}$ defined in Eq.
(\ref{ncn}) and $\langle n_h\rangle^{eq}$ in Eq. (\ref{nhn}) can be
obtained using the same approach as that described in Appendix
\ref{tip} for the derivation of Eq. (\ref{na2}) to arrive at
($\hbar\equiv2$)
\begin{equation}
\langle n_c\rangle^{eq}=-\frac{1}2\tanh(\beta_c\omega_c), \langle
n_h\rangle^{eq}=-\frac{1}2\tanh(\beta_h\omega_h), \label{ncnh}
\end{equation}
which is achieved in quasi-static limit ($x, y \rightarrow\infty$)
when $\Delta_{c,h}\rightarrow0$.  While for the finite-time
system-bath interaction interval the system is away from the thermal
equilibrium,  the populations $\langle n_C\rangle$ and $\langle
n_A\rangle$ approach the thermal values $\langle n_c\rangle^{eq}$
and $\langle n_h\rangle^{eq}$, respectively, when these intervals go
to the infinite long time limit.
  Therefore,  $\Delta_c$ and $\Delta_h$ indicate  how far the two isochoric
processes deviates from the quasistatic limit.

 From Eqs. (\ref{pw0a}) and (\ref{pqpk}), the
average heat injection, $\langle q_{c}\rangle =\int\int
q_{c}p(q_c|w_{hc})p(w_{hc})dw_{hc}dq_{c}$ can be obtained as
$(\hbar\equiv2)$
\begin{equation}
\langle q_{c}\rangle =2\omega_{c}\left[\langle
n_c\rangle+\Delta_{c}+\left(2\xi-1\right)\left(\langle
n_h\rangle+\Delta_{h}\right)\right], \label{qchh}
\end{equation}
Integrals over the distribution function $p\left(q_c|w_{hc}\right)
p\left(w_{hc}\right)$ yield the second moments of absorbed heat
$q_{c}$, $\langle q_{c}^{2}\rangle=\int\int q_{c}^{2}
\left(w_{hc}\right)p\left(q_c|w_{hc}\right)dw_{hc}dq_{c}$, which
reads $ \langle q_{c}^{2}\rangle
=8\omega_{c}^{2}\left[{1}/{4}+\left(2\xi-1\right)\left(\langle
n_h\rangle+\Delta_{h}\right) \left(\langle
n_c\rangle+\Delta_{c}\right)\right].  $ The variance of absorbed
heat $q_{c}$, $\delta q_{c}^2=\langle q_{c}^2\rangle-\langle
q_{c}\rangle^2$, then becomes
\begin{equation}
\delta{q_{c}^{2}}=4\omega_{c}^{2}\left[\frac{1}{2}-\left(2\xi-1\right)^2\left(\langle
n_h\rangle+\Delta_{h}\right)^2- \left(\langle
n_c\rangle+\Delta_{c}\right)^2\right].    \label{deqc}
\end{equation}
These variances are upper limited by the value of $2\omega_c^2$ and
they become $
\delta{q_{c}^{2}}=4\omega_{c}^{2}\left[\frac{1}{2}-\langle(
n_h\rangle)^2-\left(\langle n_c\rangle^{eq}\right)^2\right] $ for
the quasistatic cycle.

\begin{figure}[tb]
\includegraphics[width=2.8in]{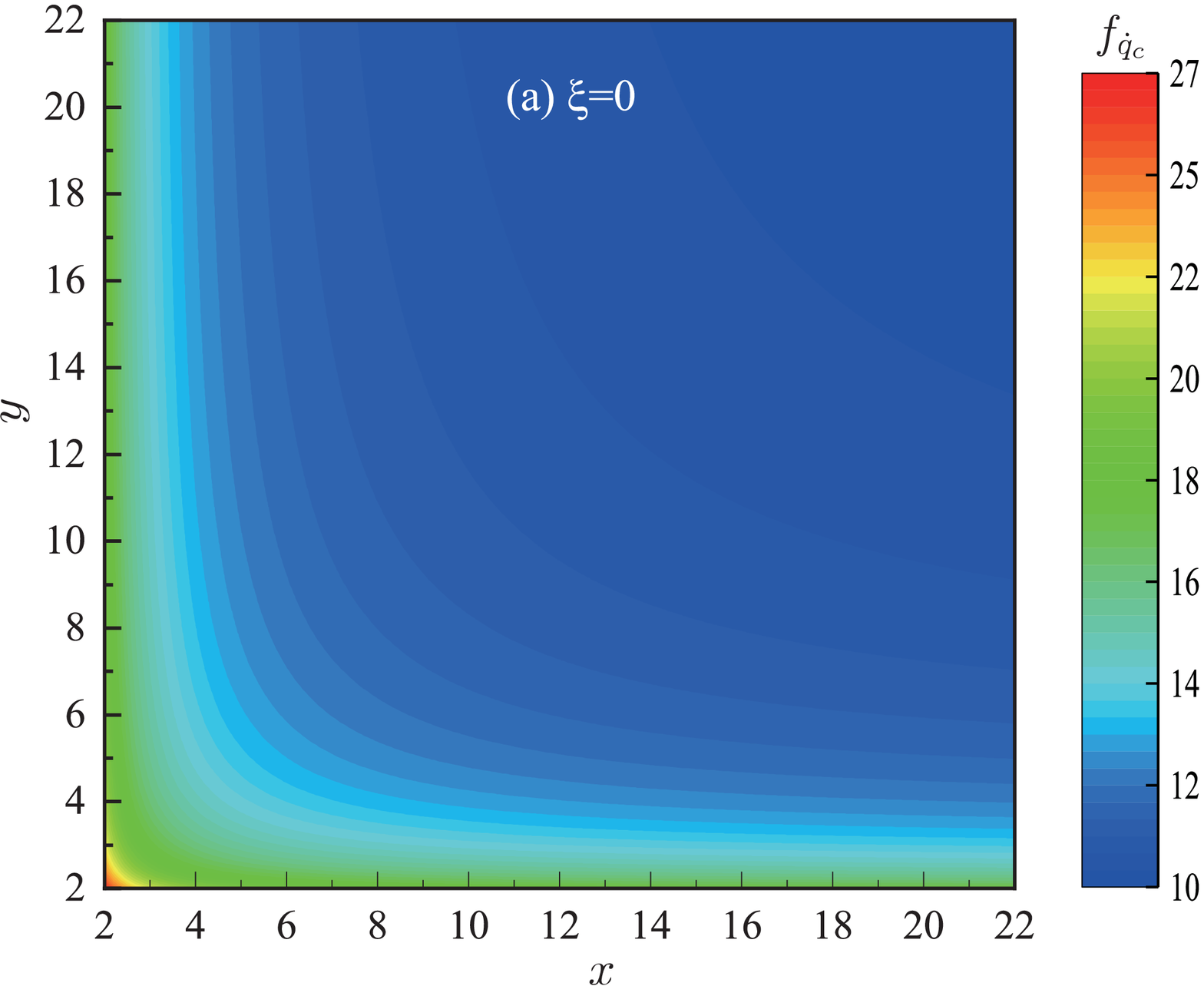}
\includegraphics[width=2.8in]{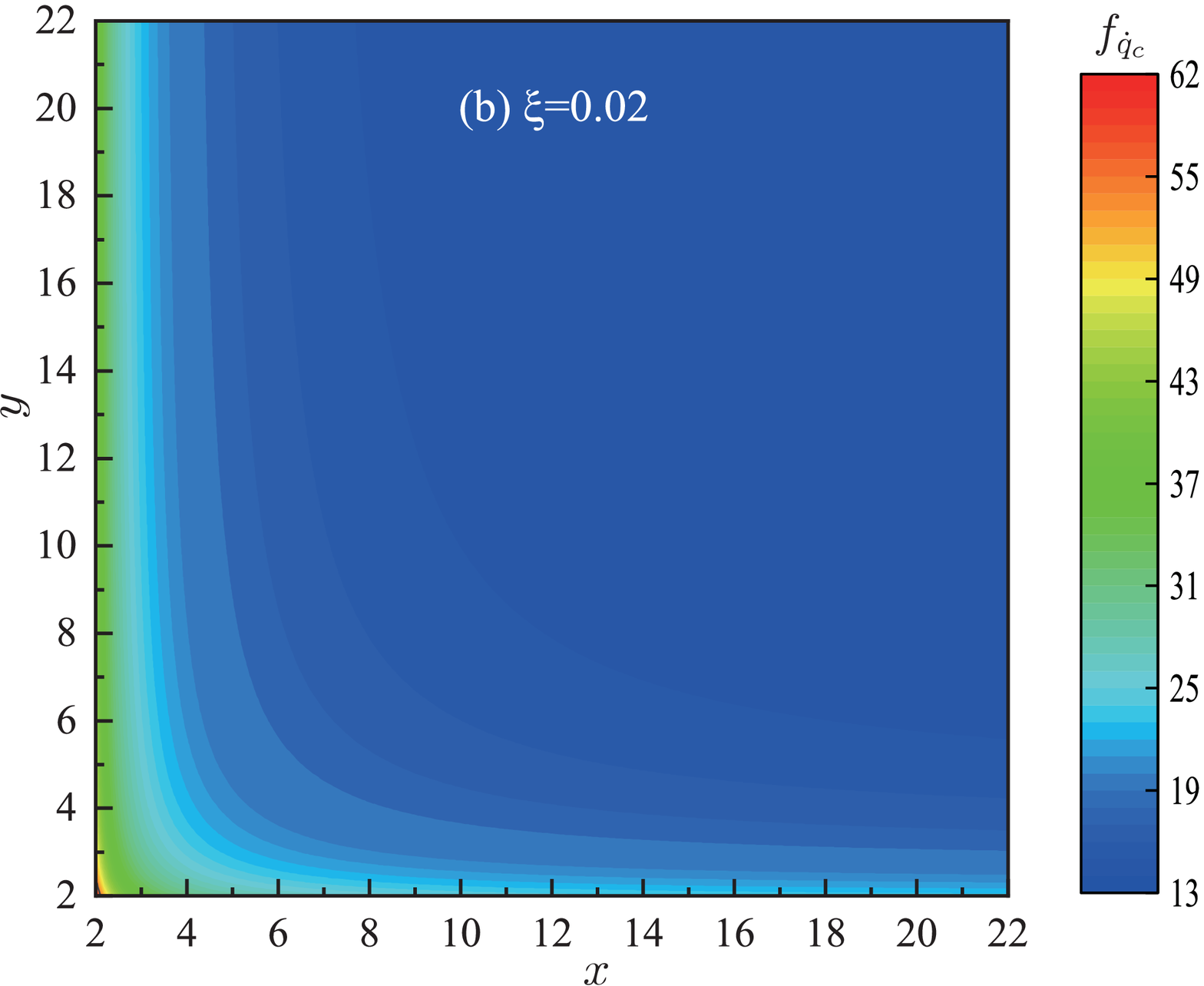}
 \caption{Contour plots of relative heat variance,  $f_{\dot{q}_c}$  in the
effective time duration $(x, y )$ plane for an adiabatic ($a$) and
nonadiabatic ($b$) driving, with $\xi=0$ and $\xi=0.02$,
respectively.
  The values of the parameters are $\beta_h=0.8$, $\beta_c=1$, $\omega_c=0.6$, and $\omega_h=1$. }\label{flqc}
\end{figure}

As $\langle\dot{q}_c\rangle=\langle q_c\rangle/\tau_{cycle}$ and
$\delta \dot{q}_c=\langle q_c\rangle/\tau_{cycle}$,  the relative
variance of cooling rare $f_{\dot{q}_c}$ can be obtained by using
Eqs. (\ref{qchh}) and (\ref{deqc}) to arrive at
\begin{equation}
f_{\dot{q}_{c}} =\frac{\delta{q_{c}}}{\langle q_{c}\rangle}
=\frac{\left[\frac{1}{2}-\left(2\xi-1\right)^2\left(\langle
n_h\rangle^{eq}+\Delta_{h}\right)^2- \left(\langle
n_c\rangle+\Delta_{c}\right)^2\right]^{\frac{1}2}}
{\left(2\xi-1\right)\left(\langle
n_h\rangle+\Delta_{h}\right)+\langle n_c\rangle^{eq}+\Delta_{c}}.
\label{fqcc}
\end{equation}
This is a monotonically decreasing function of time durations ($x$
and $y$) of system-bath interaction intervals, both for adiabatic
and nonadiabatic driving [see Figs. \ref{flqc}(a) and
\ref{flqc}(b)].
If the isochoric hot (or cold) branch is completed in an finite time
$\tau_c$ (or $\tau_h$), with finite value of $x_c$ (or $x_h$),  this
isochoric step is out of equilibrium and the fluctuations are
inevitably avoidable. In order to decrease the relative
fluctuations, we thus need to slow down the machine, which, however,
must make the average cooling rate $\langle \dot{q}_c\rangle$ down.
Comparison between \ref{flqc}(a) and \ref{flqc}(b) also shows that
the relative fluctuations are larger in nonadiabatic driving with
finite time ($\tau_{hc}$ or $\tau_{ch}$) than in adiabatic,
quasistatic evolution.

 Since no work is produced in the two isochoric
processes, the average total work per cycle is   $\langle
w_{hc}\rangle+\langle w_{ch}\rangle$, where  $\langle
w_{hc}\rangle=\int w_{hc}p(w_{hc})dw_{hc}$ and  $\langle
w_{ch}\rangle=\int w_{ch} p(w_{ch},q_c,w_{hc})$$
dw_{hc}dq_cdw_{ch}$. It follows, using Eqs. (\ref{pw0a}) and
(\ref{pwi2}),  that the total work  can be obtained,
\begin{equation}
\begin{split}
\langle w\rangle=2\left(\omega_{h}-\omega_{c}\right)\left[
\left(\langle n_c\rangle^{eq}+\Delta_{c}\right)-\left(\langle
n_h\rangle+\Delta_{h}\right)\right]
-4\xi\left[\omega_{c}\left(\langle n_h\rangle^{eq}+\Delta_{h}\right)
+\omega_{h}\left(\langle n_c\rangle^{eq}+\Delta_{c}\right)\right]
\label{w}.
\end{split}
\end{equation}
The thermodynamic coefficient of performance, defined by
$\varepsilon_{th}=\langle q_c\rangle/\langle w\rangle$, is the given
by
\begin{equation}
\varepsilon_{th}=\frac{\omega_{c}}{\omega_{h}\frac{1-\xi
\mathcal{F}}{1+\xi\mathcal{G}}-\omega_c}, \label{eth}
\end{equation}
where $\mathcal{F}=2(\langle n_c\rangle^{eq}+\Delta_c)/[(\langle
n_c\rangle+\Delta_c)-(\langle n_h\rangle^{eq}+\Delta_h)]$ and
$\mathcal{G}=2(\langle n_h\rangle^{eq}+\Delta_h)/[(\langle
n_c\rangle+\Delta_c)-(\langle n_h\rangle^{eq}+\Delta_h)]$. As
$\mathcal{F,G}<0$ and $\xi\ge0$,  the thermodynamic coefficient of
performance $\varepsilon_{th}$ increases as the adiabacity parameter
$\xi$ decreases, and it reaches its upper bound
$\varepsilon_{th}^{ad}=\omega_c/(\omega_h-\omega_c)$  in the ideal
adiabatic case when $\xi=0$. The fact that the additional heat is
dissipated into the hot reservoir due to finite time realization of
the compression or expansion, so that the additional work is input
to overcome such heat loss, suggests that cycles consisting of
nonadiabatic transformation along the expression and compression
runs less efficiently than those with ideal adiabatic strokes. While
the stochastic COP is defined by $\varepsilon=q_c/(w_{hc}+w_{ch})$,
its  probability distribution $p(\varepsilon)$ can be determined  by
$p(\varepsilon)=\int\int\int dw_{hc}d{q_c}d{w_{hc}}
p(w_{ch},q_c,w_{hc})\delta\left(\varepsilon-\frac{q_c}{w_{hc}+w_{ch}}\right)
$ to arrive at
\begin{equation}
\begin{split}
p\left(\varepsilon\right)=&2\left\{\left[\frac{1}{4}+\left(\langle
n_h\rangle+\Delta_{h}\right) \left(\langle
n_c\rangle+\Delta_{c}\right)\right]\left(1-\xi\right)^{2}
+\left[\frac{1}{4}-\left(\langle n_h\rangle^{eq}+\Delta_{h}\right)
\left(\langle n_c\rangle^{eq}+\Delta_{c}\right)\right]
\xi^{2}\right\}\delta\left(\varepsilon\right)\\
&+2\left[\frac{1}{4}+\left(\langle n_h\rangle^{eq}+\Delta_{h}\right)
\left(\langle
n_c\rangle+\Delta_{c}\right)\right]\xi^{2}\delta\left(\varepsilon+
\frac{\varepsilon_{th}^{ad}}{2\varepsilon_{th}^{ad}+1}\right)\\
&+2\left[\frac{1}{4}-\left(\langle n_h\rangle^{eq}+\Delta_{h}\right)
\left(\langle n_c\rangle^{eq}+\Delta_{c}\right)\right]
\left(1-\xi\right)^{2}\delta\left(\varepsilon-\varepsilon_{th}^{ad}\right)\\
&+\left(1-\xi\right)\xi\left[\delta\left(\varepsilon+1\right)+\delta\left(\varepsilon\right)\right]
\label{pvon}
\end{split}
\end{equation}

\begin{figure}[tb]
\includegraphics[width=3.2in]{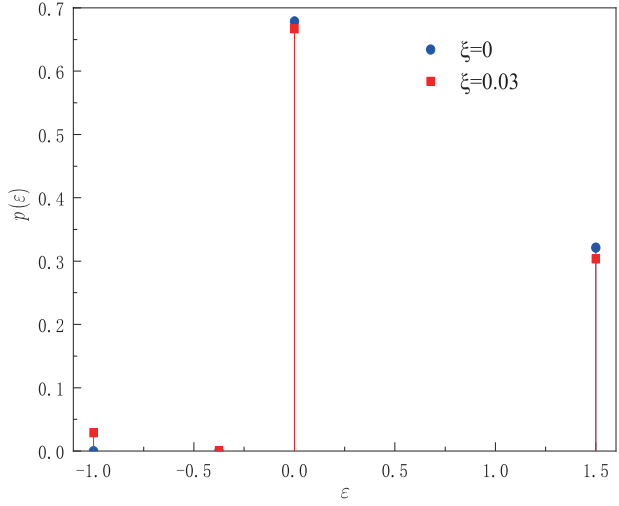}
 \caption{The  probability distribution $p(\varepsilon)$  of
the   quantum stochastic COP  for both adiabatic (blue dots) and
nonadiabatic (red squares) driving. We observe the appearance of
peaks at negative COP in the nonadiabatic case.  }\label{pepsilon}
\end{figure}
We examine the  statistics of stochastic COP  in Fig. \ref{pepsilon}
at different $\xi$ for given time durations allocated to the two
isochoric strokes ($x$ and $y$). The statistics of COP  depends on
the adiabacity parameter $\xi$ determined merely by the driving time
($\tau_{hc}$ or $\tau_{ch}$). For adiabatic driving with $\xi=0$
(blue squares),  the stochastic COP may be zero or equal to   the
adiabatic value $\varepsilon_{th}^{ad}$, with the largest peak at
zero and the second largest at $\varepsilon_{th}^{ad}$.  By
contrast, for nonadiabatic driving with $\xi>0$ (red dots),  the
negative values
[$-{\varepsilon_{th}^{ad}}/{(2\varepsilon_{th}^{ad}+1)}$ and $-1$]
of $p(\varepsilon)$ are visible due to  quantum determinacy, in
addition to nonnegative ones (zero and $\varepsilon_{th}^{ad}$).
Unlike in a quantum heat engine \cite{Lutz20} where the stochastic
efficiency can not be defined for $\xi>0$, for the quantum
refrigerator the average COP $\langle \varepsilon\rangle$ converges
and can thus be well defined.
\begin{figure}[tb]
\includegraphics[width=2.8in]{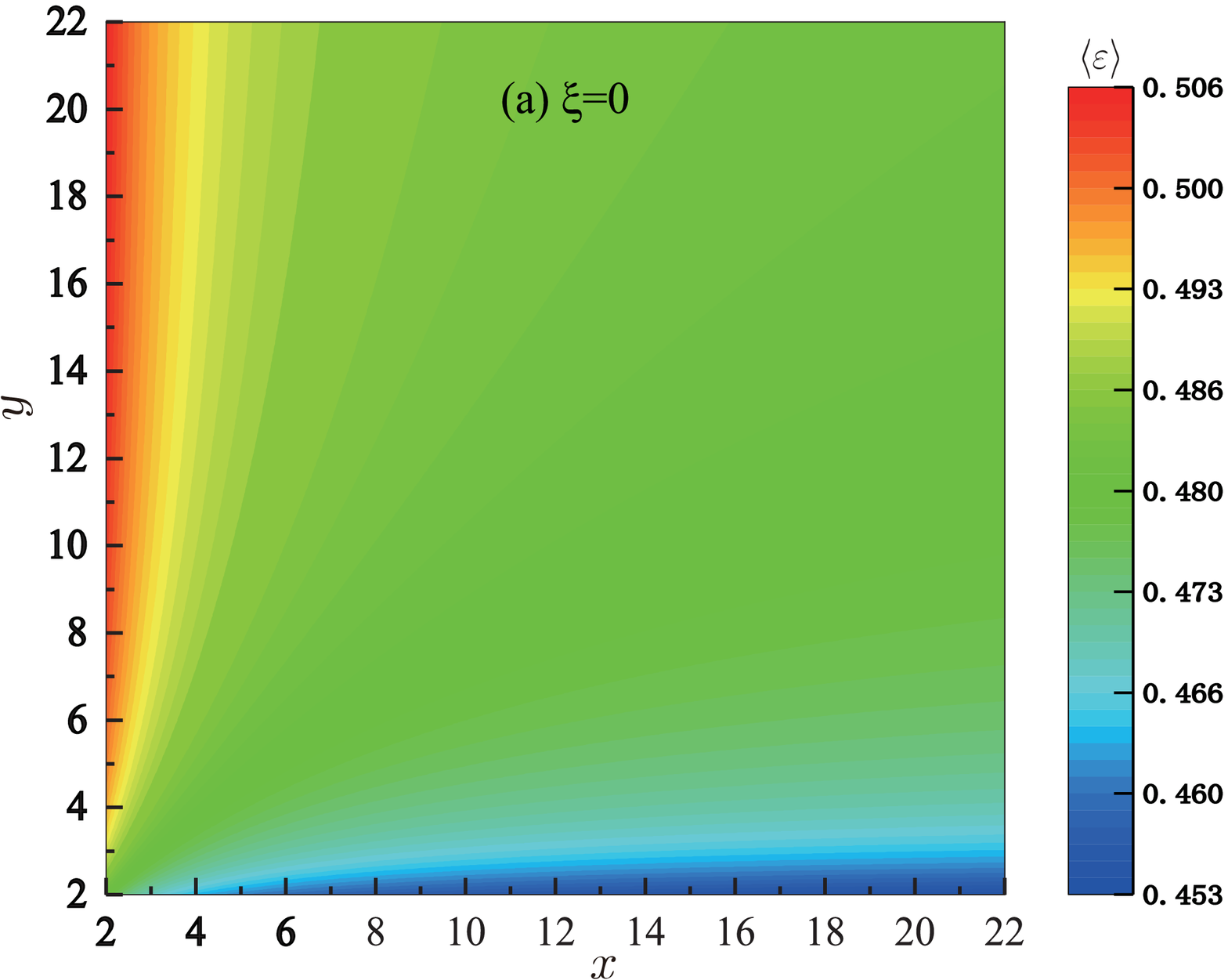}
\includegraphics[width=2.8in]{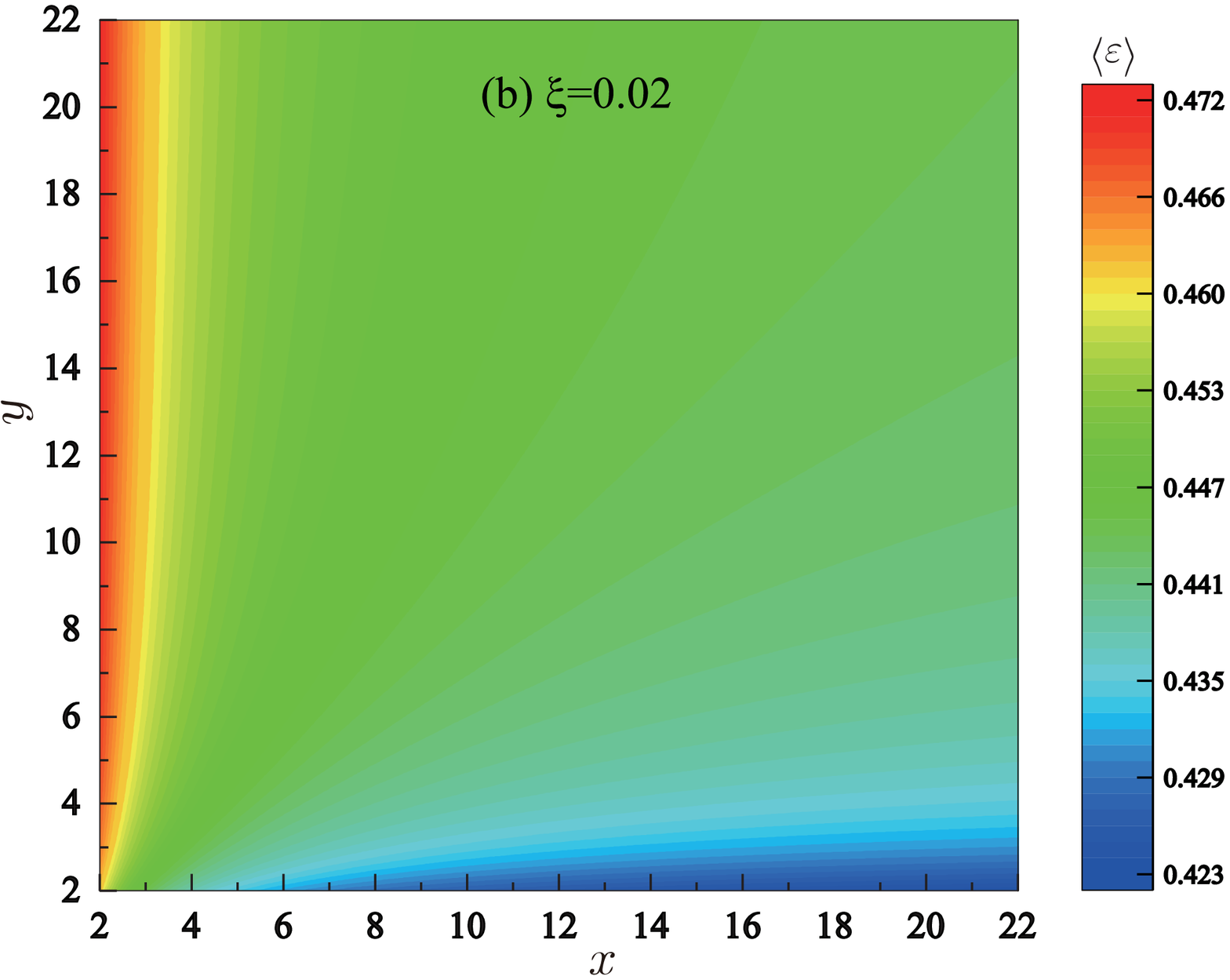}
\includegraphics[width=2.8in]{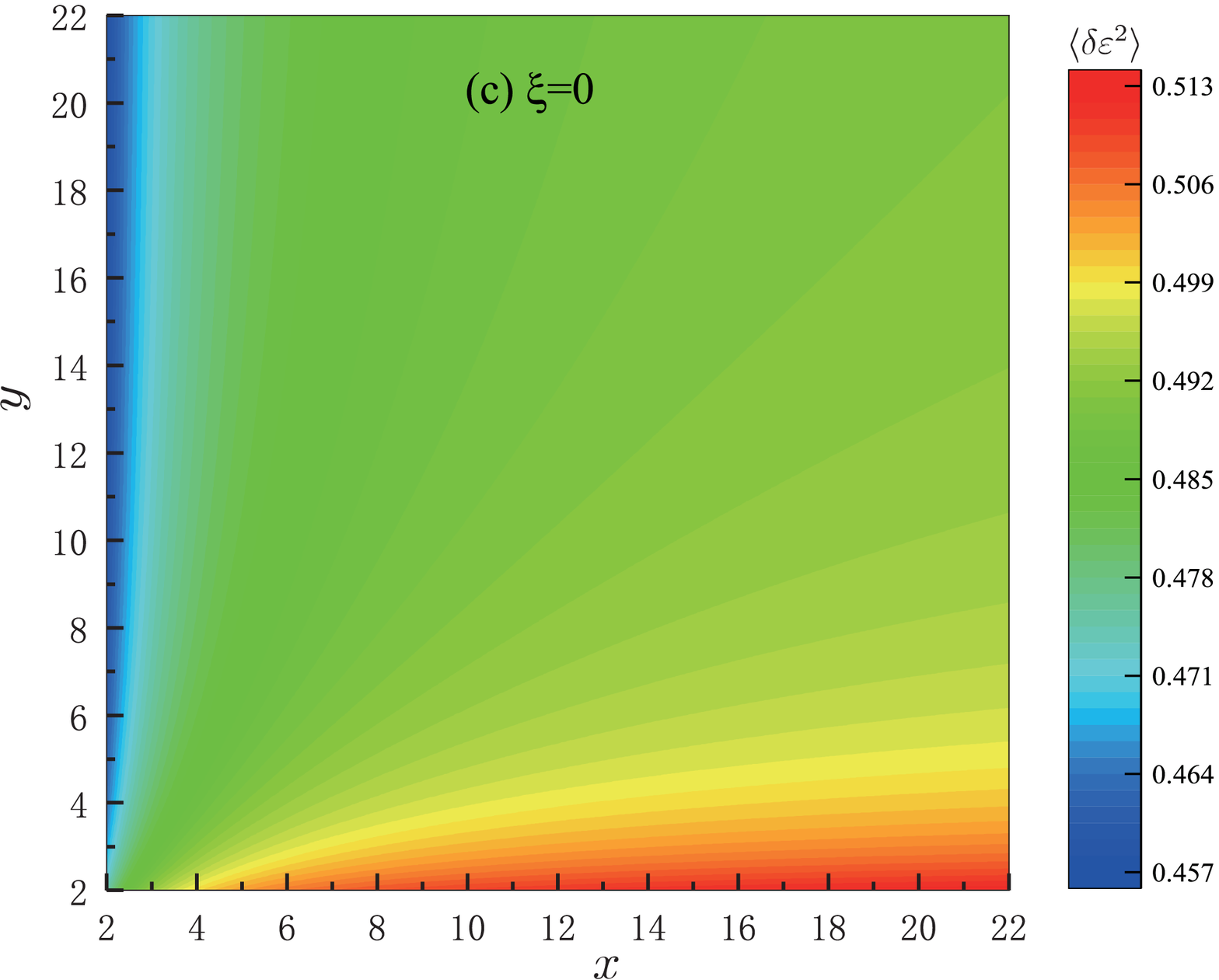}
\includegraphics[width=2.8in]{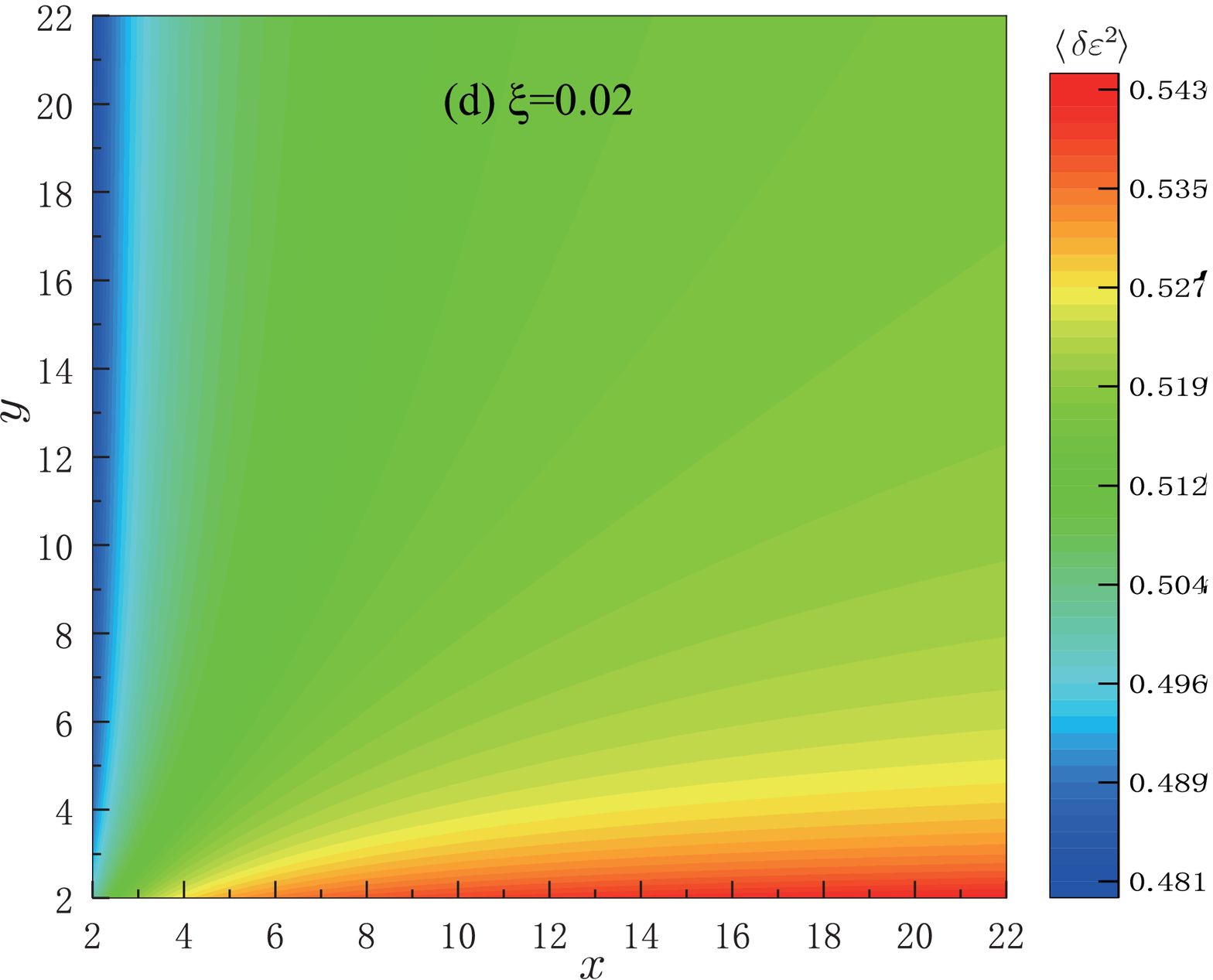}
 \caption{Contour plots of the average COP $\langle \varepsilon\rangle$ and its variance $\langle \delta\varepsilon^2\rangle$  in the
effective time duration $(x, y )$ plane for an adiabatic ($a$ and
$c$)
 and nonadiabatic ($b$ and $d$) driving, with $\xi=0$ and $\xi=0.02$, respectively.
  The values of the parameters are $\beta_h=0.8$, $\beta_c=1$, $\omega_c=0.6$, and $\omega_h=1$. }\label{mcop}
\end{figure}
Using the distribution function (\ref{pvon}),  we  find that the
first two central moments  are
\begin{equation}
\begin{split}
\langle\varepsilon\rangle=&2\left[\frac{1}{4}-\left(\langle
n_h\rangle+\Delta_{h}\right)
\left(\langle n_c\rangle^{eq}+\Delta_{c}\right)\right]\left(1-\xi\right)^{2}\varepsilon_{th}^{ad}\\
&-2\left[\frac{1}{4}+\left(\langle n_h\rangle^{eq}+\Delta_{h}\right)
\left(\langle n_c\rangle^{eq}+\Delta_{c}\right)\right]
\xi^{2}\frac{\varepsilon_{th}^{ad}}{2\varepsilon_{th}^{ad}+1}
-\left(1-\xi\right)\xi            \label{var1}
\end{split}
\end{equation}
and $ \langle\varepsilon^2\rangle\!=2\left[\frac{1}{4}-\left(\langle
n_h\rangle+\Delta_{h}\right) \left(\langle
n_c\rangle+\Delta_{c}\right)\right]\left(1-\xi\right)^{2}\left(\varepsilon_{th}^{ad}\right)^2\!
+2\left[\frac{1}{4}+\left(\langle n_h\rangle^{eq}+\Delta_{h}\right)
\left(\langle n_c\rangle^{eq}+\Delta_{c}\right)\right]
\xi^{2}\left[{\varepsilon_{th}^{ad}}/({2\varepsilon_{th}^{ad}+1})\right]^2
+\left(1-\xi\right)\xi.  $ This, combining with Eq. (\ref{var1}),
gives rise to the variance of stochastic COP,
$\delta\varepsilon^2=\langle\varepsilon^2\rangle-\langle\varepsilon\rangle^2$,
leading to
\begin{equation}
\begin{split}
\delta\varepsilon^2=&2\left[\frac{1}{4}-\left(\langle
n_h\rangle+\Delta_{h}\right)
\left(\langle n_c\rangle^{eq}+\Delta_{c}\right)\right]\left(1-\xi\right)^{2}\left(\varepsilon_{th}^{ad}\right)^2\\
&+2\left[\frac{1}{4}+\left(\langle n_h\rangle^{eq}+\Delta_{h}\right)
\left(\langle n_c\rangle^{eq}+\Delta_{c}\right)\right]
\xi^{2}\left(\frac{\varepsilon_{th}^{ad}}{2\varepsilon_{th}^{ad}+1}\right)^2\\
&-\big\{\xi\left(1-\xi\right)+\left(1-\xi\right)^2\left[2\left(\langle
n_h\rangle+\Delta_{h}\right)
\left(\langle n_c\rangle^{eq}+\Delta_{c}\right)-\frac{1}{2}\right]\varepsilon_{th}^{ad}\\
&+\xi^2\frac{\left[2\left(\langle n_h\rangle^{eq}+\Delta_{h}\right)
\left(\langle n_c\rangle^{eq}+\Delta_{c}\right)+\frac{1}{2}\right]
\varepsilon_{th}^{ad}}{2\varepsilon_{th}^{ad}+1}\big\}^{2}+\xi\left(1-\xi\right).
\label{flcop}
\end{split}
\end{equation}

 For a cycle with either adiabatic or nonadiabatic driving
branches, the average COP $\langle \varepsilon\rangle$ increases
 as time duration $y=e^{\gamma_c\tau_c}$  increases, but it decreases as time duration
 $x=e^{\gamma_h\tau_h}$ increases, see Figs. \ref{mcop}(a) and \ref{mcop}(b).
This follows from the fact that, for the machine with adiabatic or
nonadiabatic processes, the heat absorbed by the system along the
cold isochoric stroke increases as $y$ increases, and the heat
released to the hot reservoir increases as $x$ increases.   Figures
\ref{mcop}(c) and \ref{mcop}(d) show that, in contrast to $\langle
\varepsilon\rangle$, the variance $\langle \delta\varepsilon^2
\rangle$ increases as $x$ increases but decreases as $y$ increases,
thereby confirming that there is trade-off between average $\langle
\varepsilon\rangle$ and the COP fluctuations $\langle
\delta\varepsilon^2\rangle$.  We also observe that the internal
dissipation along the adiabats results in performance deterioration
for the machine by reducing the average COP $\langle
\varepsilon\rangle$  but increasing fluctuations of COP $\langle
\delta \varepsilon^2\rangle$.

\begin{figure}[tb]
\includegraphics[width=3.6in]{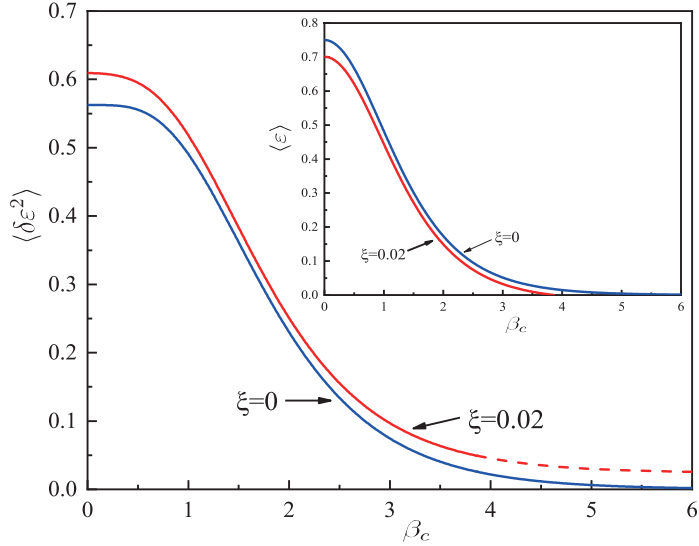}
 \caption{(Color online) The variance of stochastic COP, $\langle \delta\varepsilon^2\rangle$, and
 average COP $\langle \varepsilon\rangle$ (inset)
 as a function of the inverse  temperature of cold reservoir
 $\beta_c=0.8\beta_h$ for  $\xi=0$ (blue lines) and $\xi=0.02$ (red
 lines). The red dashed line corresponds to the variance $\langle
 \delta\varepsilon^2\rangle$ in the region of nonphysical, negative mean COP.
The parameters are $x=y=20$, $\omega_c=0.6$, and
$\omega_h=1$.}\label{flcopbc}
\end{figure}
\begin{figure}[tb]
\includegraphics[width=2.8in]{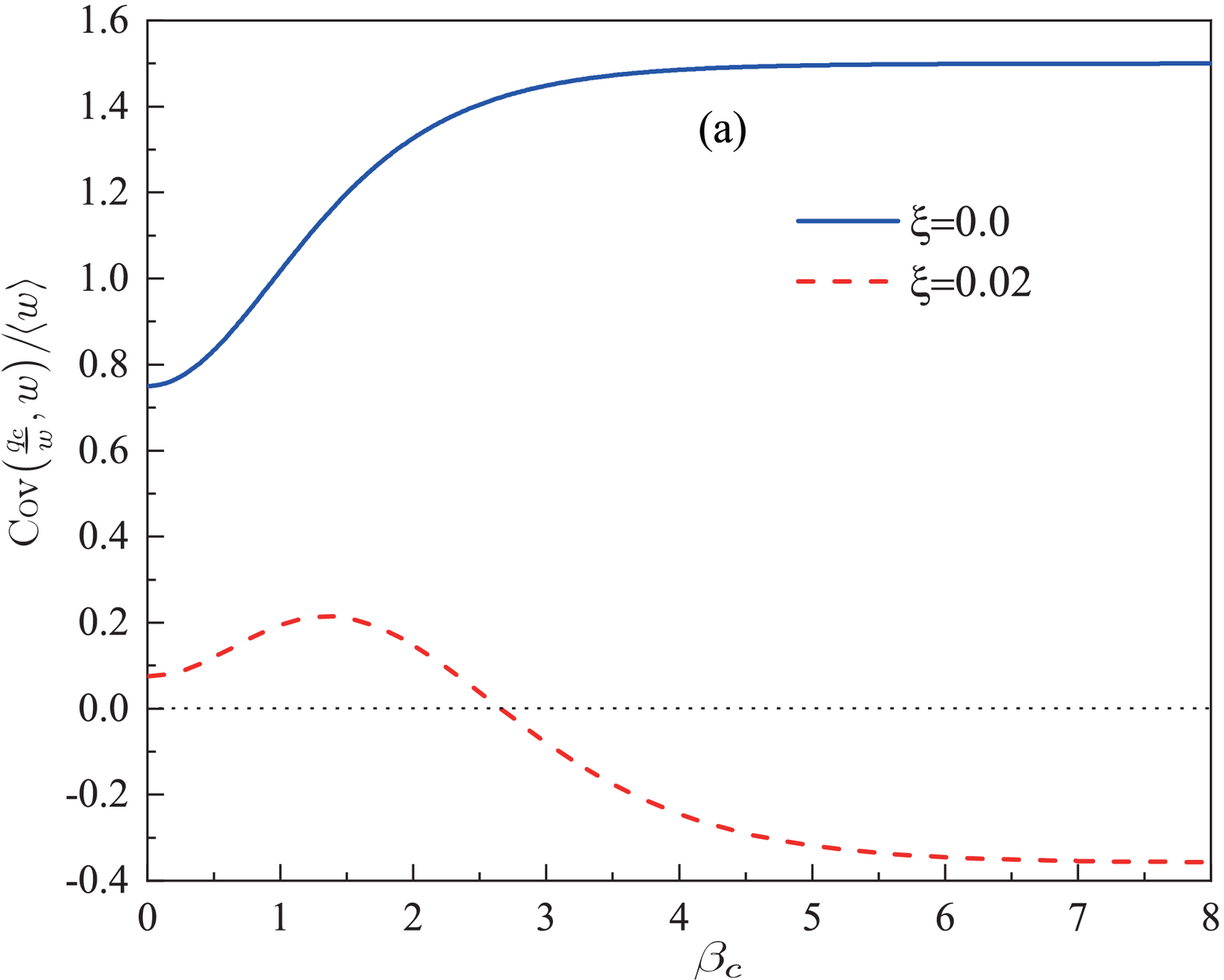}
\includegraphics[width=2.8in]{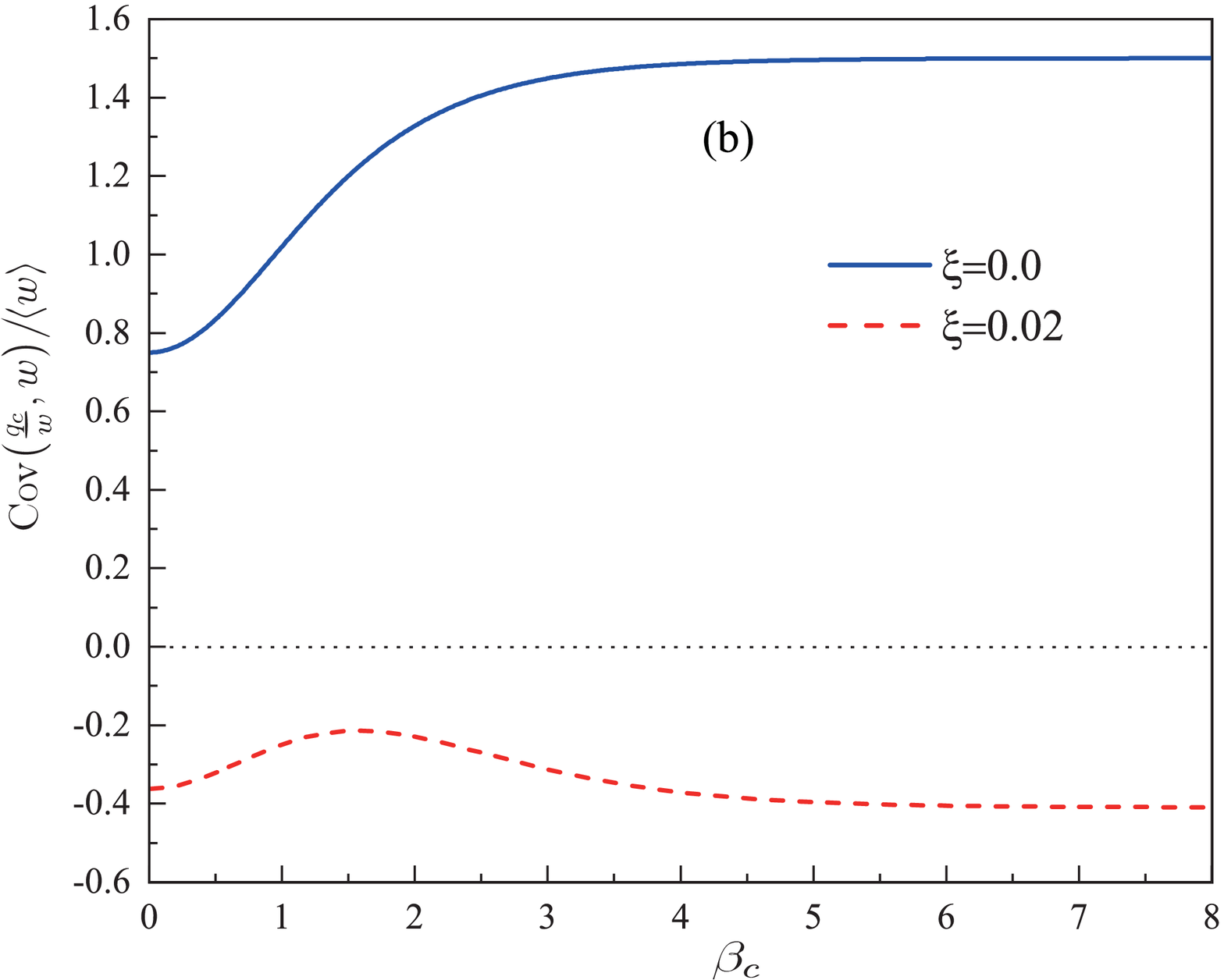}
 \caption{(Color online) The ratio ${\mathrm{Cov}\!\left(\frac{q_c}{w},w\right)\!\big/\!\langle
w\rangle}$ as a function of inverse temperature $\beta_c$
 as a function of the inverse  temperature of cold reservoir
 $\beta_c=0.8\beta_h$ for $x=y=20$ (a) and $x=y=2$ (b), where $\xi=0$ and $\xi=0.02$ are indicated by  blue lines and  red
ones, respectively. The other parameters are  $\omega_c=0.6$  and
$\omega_h=1$.}\label{ncovbc}
\end{figure}

For given time durations ($x$ and $y$) of two isochoric processes,
both the average and the variance of the stochastic COP as a
function of the inverse temperature $\beta_c$ of cold reservoir  is
shown in Fig. \ref{flcopbc}. When decreasing inverse temperature,
both $ \langle \varepsilon \rangle$ and $\langle \delta
\varepsilon^2\rangle$ grow, and as expected,  $\langle
\delta\varepsilon^2\rangle$ would vanish and $\langle
\varepsilon\rangle$ would be even negative  in the low temperature
limit. The non-positive mean COP in the low-temperature regime can
be understood that  the stochastic COP   may be negative
 due to quantum indeterminacy dominating at
low temperatures (see also Fig. \ref{pepsilon}). While the low
temperature domain is characterized by quantum fluctuations, the
high temperature region is dominated by larger thermal fluctuations.
Therefore, the variance  $\langle \delta\varepsilon^2\rangle$ gets
increased while the temperature is increasing and \emph{vice versa}.

As  the covariance between the total stochastic work $w$ and
stochastic COP $\varepsilon$ can be defined by \cite{Hei99}
\begin{equation}
\mathrm{Cov}\left(\frac{q_c}{w},w\right)=(\varepsilon_{th}-\langle \varepsilon\rangle)\langle w\rangle, \label{Cov}
\end{equation}
the  difference between the thermodynamic COP $\varepsilon_{th}$ and
average COP $\langle \varepsilon\rangle$ is determined according to
$\varepsilon_{th}-\langle
\varepsilon\rangle=\mathrm{Cov}\left(\frac{q_c}{w},w\right)/\langle
w\rangle$. The ratio
${\mathrm{Cov}\!\left(\frac{q_c}{w},w\right)\!\big/\!\langle
w\rangle}$ as a function of inverse temperature $\beta_c$ is plotted
in Figs. \ref{ncovbc}(a) and \ref{ncovbc}(b), where the time
durations along two isochores are $x=y=20$ and $x=y=2$,
respectively.  We notice that, for $\xi=0$ the ratio monotonically
increases with increasing inverse temperature. It is moreover always
positive,  for either fast or slow isochoric branch, indicating that
the thermodynamic COP $\varepsilon_{th}$ must be lager than the
average COP $\langle \varepsilon\rangle$. By contrast, this ratio
increases and then decreases as the
 temperature is lowered for $\xi>0$.  As this ratio can be either positive or negative,
  the thermodynamic COP $\varepsilon_{th}$ can be larger or smaller than the corresponding average COP $\langle\varepsilon\rangle$
  for nonadiabatic driving.

\section{Conclusion}
In summary, we have developed a general scheme allowing to determine
statistics of cooling rate and COP for a quantum Otto refrigerator
by analyzing the time evolution of the two
 isochores and two adiabats. These performance parameters as well as their statistics
are determined by the finite time durations required for completing
the two  nonadiabatic driving strokes and two isochoric branches
with incomplete thermalization. When treating an analytically
solvable two-level engine, we find that stochastic COP may be
negative due to quantum indeterminacy, and but its average value
converges and can thus be well defined. We show that there is
trade-off between these variables and their corresponding
fluctuations at zero and finite temperature, thereby indicating that
the price for enhancing the machine performance is increasing
fluctuations. We have additionally compared the average COP and the
conventional thermodynamic COP, and we found that they are positive
correlated for ideal adiabatic strokes, but their correlation may be
negative for nonadibatic branches.

\begin{appendices}
\numberwithin{equation}{section}
\section{Time evolution of population along an isochoric process } \label{tip}

When a system under external control is weakly coupled to a heat
reservoir, the quantum dynamics of the system generated by both
thermal interaction and external fields can be described by
 a semigroup
approach. The change in time of an operator $\hat{X}$ for a system
with Hamiltonian $\hat{H}$ is determined according to the master
equation \cite{Fel00, Wang19}:
\begin{equation}
\frac{d{\hat{X}}}{d t}=i[\hat{H},~\hat{X}]+\frac{\partial
{\hat{X}}}{\partial t}+\mathcal{L}_D(\hat{X}), \label{xtdx}
\end{equation}
where
\begin{equation}
\mathcal{L}_D(\hat{X})=\sum_\alpha k_\alpha
\left(\hat{V}_\alpha^\dag\left[\hat{X},
\hat{V}_\alpha^\dag\right]+\left[\hat{V}_\alpha^\dag,\hat{X}
\right]\hat{V}_\alpha\right) \label{ldx}
\end{equation}
is  the Liouville dissipative generator due to the system-reservoir
thermal interaction. $\hat{V}_\alpha$ are operators in the Hilbert
space of the system and $\hat{V}_\alpha^\dag$ are Hermitian
conjugates, and $k_\alpha$ are phenomenological positive
coefficients. In Eq. (\ref{ldx}), $[\hat{X},
\hat{V}_\alpha^\dag]=[\hat{X}, \hat{V}_\alpha^\dag]_-$ denotes
commutator for the Bose system, and  $[\hat{X},
\hat{V}_\alpha^\dag]=[\hat{a}, \hat{b}]_+$ is used for
anticommutator for the Fermi system. Substituting $\hat{X}=\hat{H}$
into Eq. (\ref{xtdx})  leads to
\begin{equation}
\left\langle\frac{d\hat{H}}{d
t}\right\rangle={\left\langle\frac{\partial{\hat{H}}}{\partial
t}\right\rangle}+\langle\mathcal{L}_D(\hat{H})\rangle.
\label{dote}
\end{equation}
This reproduces the time derivative of quantum version of the first
law of thermodynamics ${d\langle\hat{H}\rangle}/{d t}={\dbar{\langle
w\rangle}}/{d t}+{\dbar{\langle q \rangle}}/{d t}$, when the
instantaneous average power and the  average heat current are
identified as, $ {\dbar{\langle w\rangle}}/{d t} =
{\langle{\partial{\hat{H}}}/{\partial t}\rangle}$ and
${\dbar{\langle q\rangle}}/{d
t}=\langle\mathcal{L}_D(\hat{H})\rangle$, respectively.

To proceed, we choose the operators $\hat{V}^\dag$ and $\hat{V}$ as
the bosonic (fermionic) creation operator $\hat{a}^\dag$ and
annihilation operator $\hat{a}$   for the Bose (Fermi) system.
Inserting  the system Hamiltonian $\hat{H}
=\omega\hat{a}^\dag\hat{a}$ into Eq. (\ref{xtdx}) and  taking the
expectation value,  the motion of the population  $\langle
n\rangle=\langle a^\dag a\rangle$ along an isochoric process with
constant $\omega$  can be obtained as
\begin{equation}
\frac{d\langle n\rangle}{dt}
 =-\gamma(\langle n\rangle-\langle{n}\rangle^{eq}),
 \label{dn}
\end{equation}
where $\gamma=k_\downarrow-{k_\uparrow}
~(\gamma=k_\downarrow+{k_\uparrow})$ denotes the thermal
conductivity for the Bose (Fermi) system.  The detailed balance
$k_{\uparrow}=k_{\downarrow} e^{-\beta\hbar\omega}$ is assumed to be
satisfied, ensuring that the system can achieve asymptotically the
thermal state after an infinitely long system-bath interaction
duration.   Here ${\langle n
\rangle}^{eq}=\frac{1}2\frac{k_\downarrow+k_\uparrow}{k_\downarrow-{k_\uparrow}}$
$\left(\langle
n\rangle^{eq}=\frac{1}2\frac{k_\downarrow-k_\uparrow}{k_\downarrow+{k_\uparrow}}\right)$
is the asymptotic value of $\langle n(t)\rangle$ with
$t\rightarrow\infty$, and it corresponds to the equilibrium
population: $\langle{n}\rangle=1/(e^{\beta\omega}-1)$ [$\langle
n\rangle=-\frac{1}2 \tanh(\beta\omega/2)$]. From Eq. (\ref{dn}),  we
find that instantaneous population $\langle n(t) \rangle$ along the
thermalization process (staring at initial time $t=0$)  can be
written in terms of the population $\langle n(0)\rangle$,
\begin{equation}
 \langle {n}(t)\rangle=\langle n\rangle^{eq}+[\langle
 n(0)\rangle-\langle n\rangle^{eq}]e^{-\gamma t}. \label{ntgt}
\end{equation}
\section{Relation between populations at the begin and the end of a unitary driving process} \label{rep}
We consider the unitary time evolution  along the unitary adiabatic
expansion $A\rightarrow B$ to identify the explicit relation of the
populations at $A$ and $ B$. For a two level system, its
eigenenergies are $E_+=\frac{1}2\hbar\omega$, and
$E_-=-\frac{1}2\hbar\omega$. The partition function at the initial
instant $A$ along this process can be given by
\begin{equation}
Z_A=e^{-\beta_A\hbar\omega_h/2}+e^{\beta_A\hbar\omega_h/2}=2\cosh\left(\frac{\beta_A\hbar\omega_h}2\right),
\end{equation}
which, together with the occupation probabilities
$p_+=e^{-\beta\hbar\omega_h/2}/Z_A$ and
$p_-=e^{\beta\hbar\omega_h/2}/Z_A$,  gives the population at instant
$A$,
\begin{equation}
\langle
n_A\rangle=\frac{1}{2Z_A}\left(e^{-\beta\hbar\omega_h/2}-e^{\beta\hbar\omega_h/2}\right)=-\frac{1}2\tanh\left(\frac{\beta\hbar\omega_h}2\right).
 \label{na2}
\end{equation}
 The average
population at instant $B$ can then be determined according to
\begin{equation}
\begin{split}
\langle n_B\rangle&=\sum_{n,m} m p_{n\rightarrow
m}^{\tau_{hc}}p_n^0(A)\\
&=\sum_{n,m} m|\langle n|U_{\mathrm{exp}}|m\rangle|^2p_n^0(A)\\
&=\frac{1}{Z_A}\big[e^{-\beta_A\omega_h/2}\left(|\langle
+|U_{\mathrm{exp}}|+\rangle|^2-\langle
+|U_{\mathrm{exp}}|-\rangle|^2\right)\\
&+e^{\beta_A\omega_h/2}\left(|\langle
-|U_{\mathrm{exp}}|+\rangle|^2-\langle
-|U_{\mathrm{exp}}|-\rangle|^2\right)
\big]\\
&=\frac{1}{Z_A}\left[e^{-\beta_A\omega_h/2}\left(1-2\xi\right)
+e^{\beta_A\omega_h/2}\left(2\xi-1\right) \right]\\
&=\left(1-2\xi\right)\langle n_A\rangle \label{nB}
\end{split}
\end{equation}
where $\xi=|\langle \pm|U_{\mathrm{exp}}|mp\rangle|^2$ and $\langle
n_A\rangle$ was defined in Eq. (\ref{na2}). As for the two-level
system $\langle n_B\rangle <0$, $\xi$ must be upper limited by
$1/2$. Similarly, for the unitary compression $C\rightarrow D$, we
have
\begin{equation}
\langle n_D\rangle=(1-2\xi)\langle n_C\rangle.
\end{equation}
where $\xi=|\langle \pm|U_{\mathrm{com}}|\mp\rangle|^2$.

\end{appendices}

 \textbf{Acknowledgements}
This work is  supported by National Science Foundation of China
(Grant Nos. 11875034 and 11505091), and by the State Key Program  of
China under Grant NO. 2017YFA0304204. J. H. W. also acknowledges the
financial support from the Major Program of Jiangxi Provincial
Natural Science Foundation (Grant No. 20161ACB21006).

\end{document}